\documentclass[twoside,twocolumn,9pt]{article}
\usepackage{extsizes}
\usepackage[super,sort&compress,comma]{natbib} 
\usepackage[version=3]{mhchem}
\usepackage[left=1.5cm, right=1.5cm, top=1.785cm, bottom=2.0cm]{geometry}
\usepackage{balance}
\usepackage{mathptmx}
\usepackage{sectsty}
\usepackage{graphicx} 
\usepackage{lastpage}
\usepackage[format=plain,justification=justified,singlelinecheck=false,font={stretch=1.125,small,sf},labelfont=bf,labelsep=space]{caption}
\usepackage{float}
\usepackage{fancyhdr}
\usepackage{fnpos}
\usepackage[english]{babel}
\addto{\captionsenglish}{%
  
}
\usepackage{array}
\usepackage{droidsans}
\usepackage{charter}
\usepackage[T1]{fontenc}
\usepackage[usenames,dvipsnames]{xcolor}
\usepackage{setspace}
\usepackage[compact]{titlesec}
\usepackage{hyperref}

\usepackage{epstopdf}

\definecolor{cream}{RGB}{222,217,201}

\begin{document}

\pagestyle{fancy}
\thispagestyle{plain}
\fancypagestyle{plain}{
\renewcommand{\headrulewidth}{0pt}
}

\makeFNbottom
\makeatletter
\renewcommand\LARGE{\@setfontsize\LARGE{15pt}{17}}
\renewcommand\Large{\@setfontsize\Large{12pt}{14}}
\renewcommand\large{\@setfontsize\large{10pt}{12}}
\renewcommand\footnotesize{\@setfontsize\footnotesize{7pt}{10}}
\makeatother

\renewcommand{\thefootnote}{\fnsymbol{footnote}}
\renewcommand\footnoterule{\vspace*{1pt}%
\color{cream}\hrule width 3.5in height 0.4pt \color{black}\vspace*{5pt}} 
\setcounter{secnumdepth}{5}

\makeatletter 
\renewcommand\@biblabel[1]{#1}            
\renewcommand\@makefntext[1]%
{\noindent\makebox[0pt][r]{\@thefnmark\,}#1}
\makeatother 
\renewcommand{\figurename}{\small{Fig.}~}
\sectionfont{\sffamily\Large}
\subsectionfont{\normalsize}
\subsubsectionfont{\bf}
\setstretch{1.125} 
\setlength{\skip\footins}{0.8cm}
\setlength{\footnotesep}{0.25cm}
\setlength{\jot}{10pt}
\titlespacing*{\section}{0pt}{4pt}{4pt}
\titlespacing*{\subsection}{0pt}{15pt}{1pt}

\fancyfoot{}
\fancyfoot[LO,RE]{\vspace{-7.1pt}\includegraphics[height=9pt]{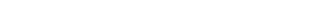}}
\fancyfoot[CO]{\vspace{-7.1pt}\hspace{13.2cm}\includegraphics{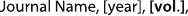}}
\fancyfoot[CE]{\vspace{-7.2pt}\hspace{-14.2cm}\includegraphics{head_foot/RF}}
\fancyfoot[RO]{\footnotesize{\sffamily{1--\pageref{LastPage} ~\textbar  \hspace{2pt}\thepage}}}
\fancyfoot[LE]{\footnotesize{\sffamily{\thepage~\textbar\hspace{3.45cm} 1--\pageref{LastPage}}}}
\fancyhead{}
\renewcommand{\headrulewidth}{0pt} 
\renewcommand{\footrulewidth}{0pt}
\setlength{\arrayrulewidth}{1pt}
\setlength{\columnsep}{6.5mm}
\setlength\bibsep{1pt}


\newcommand{\comment}[1]{{\color{black} #1}}

\makeatletter 
\newlength{\figrulesep} 
\setlength{\figrulesep}{0.5\textfloatsep} 

\newcommand{\topfigrule}{\vspace*{-1pt}%
\noindent{\color{cream}\rule[-\figrulesep]{\columnwidth}{1.5pt}} }

\newcommand{\botfigrule}{\vspace*{-2pt}%
\noindent{\color{cream}\rule[\figrulesep]{\columnwidth}{1.5pt}} }

\newcommand{\dblfigrule}{\vspace*{-1pt}%
\noindent{\color{cream}\rule[-\figrulesep]{\textwidth}{1.5pt}} }

\makeatother

\twocolumn[
  \begin{@twocolumnfalse}
{\includegraphics[height=30pt]{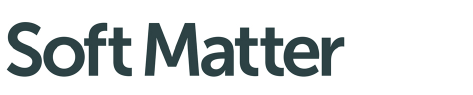}\hfill\raisebox{0pt}[0pt][0pt]{\includegraphics[height=55pt]{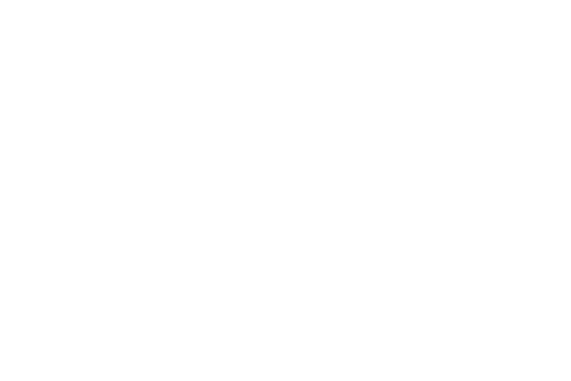}}\\[1ex]
\includegraphics[width=18.5cm]{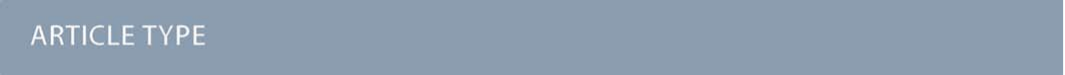}}\par
\vspace{1em}
\sffamily
\begin{tabular}{m{4.5cm} p{13.5cm} }

\includegraphics{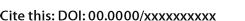} & \noindent\LARGE{\textbf{Restructuring a passive colloidal suspension using a rotationally driven particle\textit{$^{\dag}$}}} \\
\vspace{0.3cm} & \vspace{0.3cm} \\

 & \noindent\large{Shih-Yuan Chen,\textit{$^{a\ddag}$} Hector Lopez-Rios,\textit{$^{b\ddag}$} Monica Olvera de la Cruz,\textit{$^{a,b\ast}$} and Michelle Driscoll\textit{$^{a\ast}$}} \\

\includegraphics{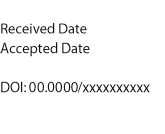} & \noindent\normalsize{\comment{The interaction between passive and active/driven particles has introduced a new way to control colloidal suspension properties from particle aggregation to crystallization. Here, we focus on the hydrodynamic interaction between a single rotational driven particle and a suspension of passive particles near the floor. Using experiments and Stokesian dynamics simulations that account for near-field lubrication, we demonstrate that the flow induced by the driven particle can induce long-ranged rearrangement in a passive suspension. We observe an accumulation of passive particles in front of the driven particle and a depletion of passive particles behind the driven particle. This restructuring generates a pattern that can span a range more than $10$ times of the driven particle’s radius. We further show that size scale of the pattern is only a function of the particle’s height above the floor.}}
\end{tabular}

 \end{@twocolumnfalse} \vspace{0.6cm}

  ]

\renewcommand*\rmdefault{bch}\normalfont\upshape
\rmfamily
\section*{}
\vspace{-1cm}


\footnotetext{\textit{$^{a}$~Department of Physics \& Astronomy, Northwestern University, Evanston, Illinois, 60208, USA.}}
\footnotetext{\textit{$^{b}$~Department of Materials Science \& Engineering, Northwestern University, Evanston, Illinois, 60208, USA.}}
\footnotetext{\textit{$^{\ast}$~Corresponding authors: m-olvera@northwestern.edu; michelle.driscoll@northwestern.edu}}
\footnotetext{\textit{$^{\ddag}$~These authors contributed equally.}}
\footnotetext{\dag~Electronic Supplementary Information (ESI) available: [details of any supplementary information available should be included here]. See DOI: 10.1039/cXsm00000x/}



\section{Introduction}

Colloids have been extensively used to explore the relation between structure and function of materials. Their structure is easily observable with a simple optical microscope \cite{li_colloidal_2011,zhang_toward_2015,manoharan_colloidal_2015}, and colloidal particles can be actuated to roll \cite{driscoll_unstable_2017}, spin \cite{sabrina_shape-directed_2018}, and oscillate \cite{zhang_quincke_2021}.
Moreover, one can easily tune particle interactions in a colloidal system using straightforward modifications to particle size, shape, and surface chemistry, which allows for the design of exotic material properties such as tunable shear-jamming \cite{chen_leveraging_2023} and patterned wettability \cite{shao_superwettable_2019}.

Generally, self-assembly is driven by interparticle forces and entropic forces. External applied fields, however, unlock self-assembly pathways that are otherwise inaccessible including avalanches \cite{driscoll_leveraging_2019} and configurations that encode memory \cite{kaz_physical_2012,zhang_polar_2022}; indeed, the soft matter community has established that self-propelling particles can tune self-assembly \cite{massana-cid_active_2018,omar_swimming_2019}. Additional degrees of freedom, such as in mixtures of active and passive particles, enhance the phase space of microstructures \cite{superlatts_hydro,mallory_active_2018,madden_hydrodynamically_2022}. The motion of self-propelling particles has been shown to enhance the diffusivity of passive particles \cite{mino_enhanced_2011,jepson_enhanced_2013}, create clusters of passive particles around self-propelling particles \cite{palacci_living_2013,katuri_inferring_2021}, or induce phase separation between particle types \cite{mccandlish_spontaneous_2012,cates_motility-induced_2015,stenhammar_activity-induced_2015,wysocki_propagating_2016,smrek_small_2017,dolai_phase_2018}. 
Moreover, as the self-propelling particles exert forces on their surroundings, they continue reconfiguring the local structure already built \cite{katuri_inferring_2021,singh_interaction_2022}.
When active particles are added to an equilibrium passive crystal structure, their activity accelerates the annealing process, leading to large-scale single crystals \cite{ramananarivo_activity-controlled_2019}. 
%
%
Through active-passive interactions, active particles via self-diffusiophoresis \cite{illien_fuelled_2017} and driven particles via external fields have the potential to tune the aggregation of passive particles \cite{massana-cid_active_2018,omar_swimming_2019}, and ship cargo passive particles to desired locations \cite{petit_selective_2012,zhang_targeted_2012}. 
These examples demonstrate the complexity of active-passive and driven-passive interactions. To understand how these particles reshape material structure, it is crucial to understand how they interact with their surroundings in a colloidal suspension. In many cases, hydrodynamics plays a significant role to explain the restructuring and the emergent patterns \cite{katuri_inferring_2021,marchetti_hydrodynamics_2013,boniface_hydrodynamics_2023}.

\begin{figure*}
\centering
\includegraphics[width=.9\linewidth]{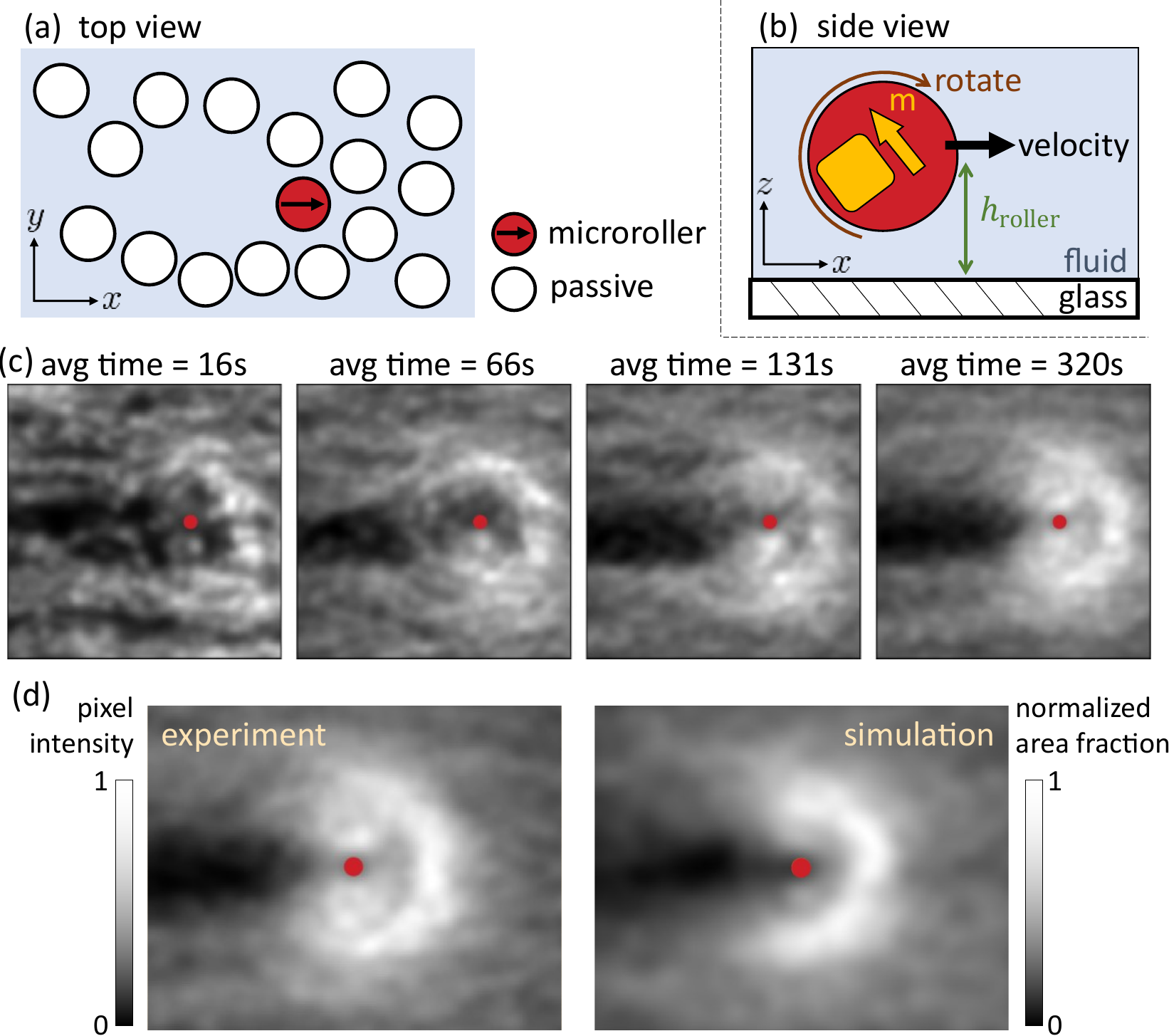}
\caption{
\textbf{Microrollers alter microstructure in passive suspensions.} We dope driven particles (microrollers) in a passive colloidal suspension, as demonstrated in the schematic (a) A small quantity of microrollers (driven particles) are added to a suspension of passive particles. (b) Microrollers contain a permenant magnetic moment, m, and are acuated by applying a uniform rotating magnetic field.  This actuation generates strong advective flows, which scale with $h_{\mathbf{roller}}$, the height of the particle above the surface; these flows are the driver for restructuring the passive suspension. (c) Microrollers restructure the passive suspension by modulating the mean local  density; this resructuring becomes more and more apparent as we average over longer and longer times. (d) Restructuring of the passive suspensions results in the emergence of a steady-state pattern. Left image is the experimental result, in which brighter areas indicate a higher local intensity, which is correlated to a higher concentration of passive particles.  Right image shows the result of Stokesian dynamics simulations, which reproduce the same pattern seen in the experiments.
}
\label{fig:roller}
\end{figure*}

\begin{figure*}
\centering
\includegraphics[width=.8\linewidth]{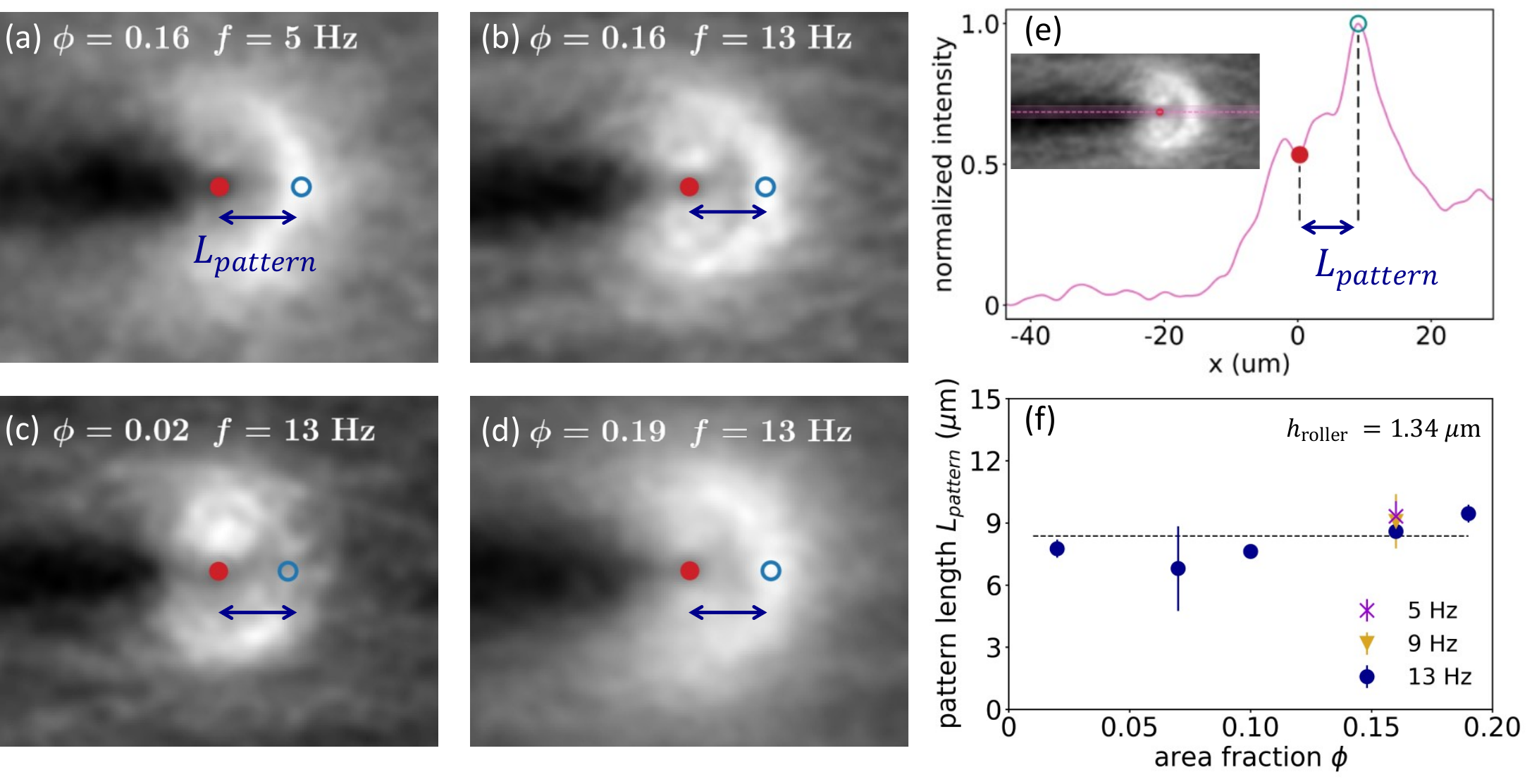}
\caption{\textbf{The length of the new structure is independent of the microroller's velocity (the rotatinal frequency) and the area fraction of the passive particles.} 
(a-b) Pattern length at two different actuation frequencies: (a) 5 Hz and (b) 13 Hz.  We find that the pattern size, $L_{\mathrm{pattern}}$, is independent of microroller velocity.  (c-d) Similarly, altering the mean area fraction of passive particles has no effect on $L_{\mathrm{pattern}}$.  (c) $\phi$ = 0.02 (d) $\phi$ = 0.19.
(e) To quantify $L_{\mathrm{pattern}}$, we draw a box across the microroller along the $x$ axis of the pattern and measure the average intensity in the y axis, as illustrated in the inset; the pink dashed line indicates the center of the box. Then, we find the peak of the intensity (the blue open circle) in front of the roller (the orange circle), and measure the distance from the peak to the microroller (the blue double arrow). 
(f) $L_{\mathrm{pattern}}$ is independent of both microroller velocity (actuation frequency $f$) and $\phi$. We find the average $L_{\mathrm{pattern}}$ is $8.9 \pm 0.9 \ \mu$m (the black dashed line in (f)). 
}
\label{fig:robust}
\end{figure*}

In this work, we examine how a driven particle, called a microroller \cite{delmotte_hydrodynamic_2017,rollers_stability_analysis}, immersed in a passive particle suspension can restructure its surroundings through experiments and Stokesian dynamics simulations \comment{with lubrication corrections (see Materials section).}
Here a microroller rearranges the passive particles through hydrodynamic interactions, creating an accumulation zone around itself and a depletion tail in its wake. We observe that a steady-state pattern emerges from these interactions, and this pattern is $10$ times larger than the microroller radius and is three dimensional in nature. We note that in striking contrast to restructuring created by actively dragging a particle through a suspension \cite{zia_active_2018}, the highest concentration of passive particles in the accumulation area is roughly $10$ microroller radii away from the microroller location. 
This underscores the mechanism of restructuring as the long-range flow field generated by the microroller. Due to the sensitive dependence of hydrodynamic interactions on the particle's height above the chamber floor, we find that the structure of the pattern is modified by changing the height of either the driven or passive particles.

\section{Results}

We experimentally study suspensions of passive particles in water doped with magnetically driven particles, named the microrollers, see Fig.~\ref{fig:roller}(a). 
Both the microrollers and passive particles are denser than water, so they sediment to the floor of the suspension's container, forming a quasi-2D layer of particles. Both types of the particles share the same size ($2 \ \mu$m in diameter). Due to the thermal fluctuation and electrostatic repulsion, both types of particles do not contact the floor but instead hover above the floor surface at an equilibrium height, $h_{\mathrm{roller}}$ and $h_{\mathrm{passive}}$ respectively, as shown in Fig.~\ref{fig:roller}(b). 
The microroller's translational velocity is determined by two parameters, the microroller's height $h_\mathrm{roller}$ and its rotational frequency $f$. 
We select the rotational frequency to be between $5$~Hz to $13$~Hz in our experiments so that the speed of the microroller is linearly proportional to the rotating frequency (see SI.1). 

In the driven-passive colloidal mixture, passive particles are initially distributed randomly around the microroller in the $xy$ plane. 
When the external rotating magnetic field is on, the microroller is translating, and the flows generated by the microroller redistribute the passive particles (see Movie S1). 
We observe that the passive particles are restructured as follows: there is an accumulation of passive particles in the region of the direction of motion of the microroller (in the $+\hat{x}$ direction), while there is a depleted region opposite to the direction of translation (in the $-\hat{x}$ direction). 
We note that the average microroller speed is constant (see SI.2); the system is in a non-equilibrium steady state.

To quantify the restructuring, we calculate the time averaged number density of passive particles $\left < \rho_{\mathrm{passive}}\left( \mathbf{r} \right) \right >_t$ in the microroller's frame of reference.
In the microroller's frame, the microroller is static, and it is the passive particles that move around the microroller (see Movie S2).
A pattern in the passive particle distribution emerges around the microroller as we average more and more frames, as shown in Fig.~\ref{fig:roller}(c).
Brighter regions indicate a longer presence of passive particles while darker regions signal the opposite. The emergent pattern reveals a depletion of passive particles in the $-\hat{x}$ direction to the microroller, and a greater presence of passive particles in the vicinity of the microroller.
Surprisingly, the peak of the accumulation (the brightest location) in the pattern in the $+\hat{x}$ direction to the microroller is much larger than the particle itself, approximately $10$ times the radius of the microroller ($10 \, \mu \mathrm{m}$). 

To complement our experimental work, we study this system computationally using Stokesian dynamics with lubrication corrections (see Movie S3 for the microroller in the lab frame and S4 for the microroller in the microroller's frame). \comment{Therefore, near-field or lubrication corrections are included between particles in simulations. We also model particles as perfectly rigid spheres to account for their finite size}. Previous work has shown that a microroller rolls at constant angular velocity imposed by a rotating magnetic field rather than experiencing a constant torque \cite{sprinkle_driven_2020}. Thus, to resemble experiments we study a rotating particle with constant angular velocity in a region where we have initially fixed the area fraction of passive particles. For suspensions at finite temperature we stochastically evolve the system to solve for Brownian dynamics and correctly account for thermal fluctuations. Importantly, the height of the microroller ($h_{\mathrm{roller}}$) sets its velocity profile as a function of angular velocity. Obtaining $h_{\mathrm{roller}}$ from experiments is challenging, thus we use the velocity profile (the microroller's velocity as a function of rotational frequencies) to determine $ \left < h_{\mathrm{roller}} \right > $ to use in simulations. 

Moreover, we use Gaussian smoothing in all $\left < \rho_{\mathrm{passive}}\left( \mathbf{r} \right) \right >_t$ using a variance the size of the passive particle. This adds particle areal effects in $\left < \rho_{\mathrm{passive}}\left( \mathbf{r} \right) \right >_t$, and thus can be comparable to the experimental emergent pattern. In Fig.~\ref{fig:roller}(d), we see that the simulation reproduces the same pattern as observed in experiments, and the areas of high intensity in experiments correspond to areas of high particle density in simulations.

In order to determine what parameters control the features and the size of the emergent pattern, we carry out experiments by varying the rotating frequency $f$ from $5$~Hz to $13$~Hz, \comment{which linearly increases the microroller speed. We also perform experiments by varying} the area fraction of passive particles $\phi_\mathrm{passive}$ from $2\%$ to $19 \%$. \comment{As the temperature, particle density, and the electrostatic repulsions remain the same, $h_{\mathrm{roller}}$ is kept at $h_{\mathrm{roller}}= 1.34 \ \mu$m in all experiments.}
In no case do we observe that the pattern changes, see Fig.~\ref{fig:robust}(a-d).
To quantify our results, we compute a characteristic pattern length 
$L_{\mathrm{pattern}}$ by measuring the intensity across the pattern along the x axis and define $L_{\mathrm{pattern}}$ to be the distance between the microroller center to the intensity peak, see Fig.~\ref{fig:robust}(e). 
As expected from the experimental images, in Fig.~\ref{fig:robust}(f), $L_{\mathrm{pattern}}$ is invariant with respect to $\phi_\mathrm{passive}$ and $f$. \comment{The results demonstrate that the pattern length is independent of the microroller speed or the area fraction of passive particles at a given $h_\mathrm{roller}$.}

\begin{figure}
\centering
\includegraphics[width=.5\linewidth]{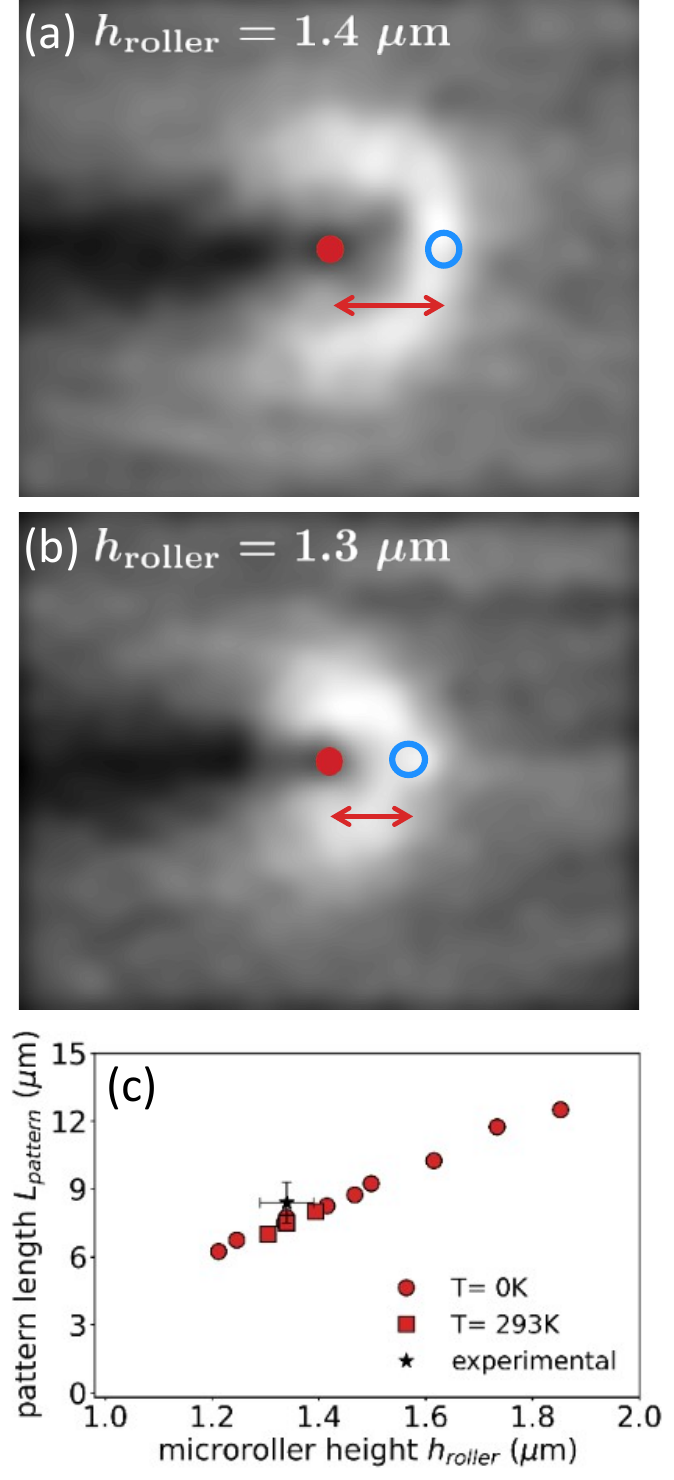}
\caption{
\textbf{The pattern length is set only by the microroller's height above the floor.} 
We perform simulations at different microroller's heights $h_{\mathrm{roller}}$ at $T=0$K and $T= 293$K to investigate $L_{\mathrm{pattern}}$ as a function of $h_{\mathrm{roller}}$. We observe that $L_{\mathrm{pattern}}$ becomes smaller as $h_{\mathrm{roller}}$ decreases: (a) $h_{\mathrm{roller}}= 1.4 \ \mu$m, (b) $h_{\mathrm{roller}}= 1.05 \ \mu$m, and is identical at $T= 0$~K and $T = 293$~K. 
}
\label{fig:notrobust}
\end{figure}
Using simulations, we also investigate the influence of changing $h_{\mathrm{roller}}$. 
The simulations are carried out at $f= 10$~Hz and $\phi = 0.17$ at $T=0$K and $T=293$K. The results are shown in Fig.~\ref{fig:notrobust}. We measure the peak area fraction of the pattern to the microroller to extract $L_{\mathrm{pattern}}$. As shown in Fig.~\ref{fig:notrobust}(c), $L_{\mathrm{pattern}}$ is directly proportional to $h_{\mathrm{roller}}$, even in the the presence of thermal fluctuations. \comment{Both the flow field generated by the microroller and the microroller speed depend on both $h_{\mathrm{roller}}$ and $f$. As shown in Fig.~\ref{fig:robust}, $L_\mathrm{pattern}$ is independent of $f$ for a given $h_{\mathrm{roller}}$. Therefore, the change of the flow field as a result of modifying $h_{\mathrm{roller}}$ must be the reason why the pattern changes. We will further elaborate the pattern mechanism in the Discussion section.}


\section{Discussion}
\label{sec:discussion}


To understand the origin of the emergent pattern from experiments we must determine the principle stresses at play in suspensions at finite temperature. Hydrodynamic forces between particles can be impacted by thermal fluctuations as they will disrupt particle's trajectories along streamlines. To estimate the relative influence of thermal fluctuations compared to the advective flows, we calculate the P\'eclet number, $\mathrm{Pe} = \frac{R_\mathrm{passive} u_\mathrm{fluid}}{D_\mathrm{passive}}$, where $R_\mathrm{passive}$ is the average passive particle radius, $u_\mathrm{fluid}$ is approximately the maximum velocity of the fluid due to a microroller, and $D_\mathrm{passive}$ is the passive particle diffusion coefficient. Using the values $R_\mathrm{passive} = 1 \ \mu \mathrm{m}$, $u_\mathrm{fluid} = 50 \ \mu\mathrm{m}/\mathrm{s}$, and $D_\mathrm{passive} = 0.15 \ \mu \mathrm{m}^2/\mathrm{s}$, we calculate that $\mathrm{Pe} = 333 \gg 1$; and thus reveals that passive particle transport is dominated by microroller-generated advective flows rather than from thermal fluctuations. Therefore, computationally less expensive simulations at $T=0 \mathrm{K}$ are sufficient to understand the origin of the pattern formation; Brownian motion does not alter the average size of the pattern. 


\begin{figure*}
\centering
\includegraphics{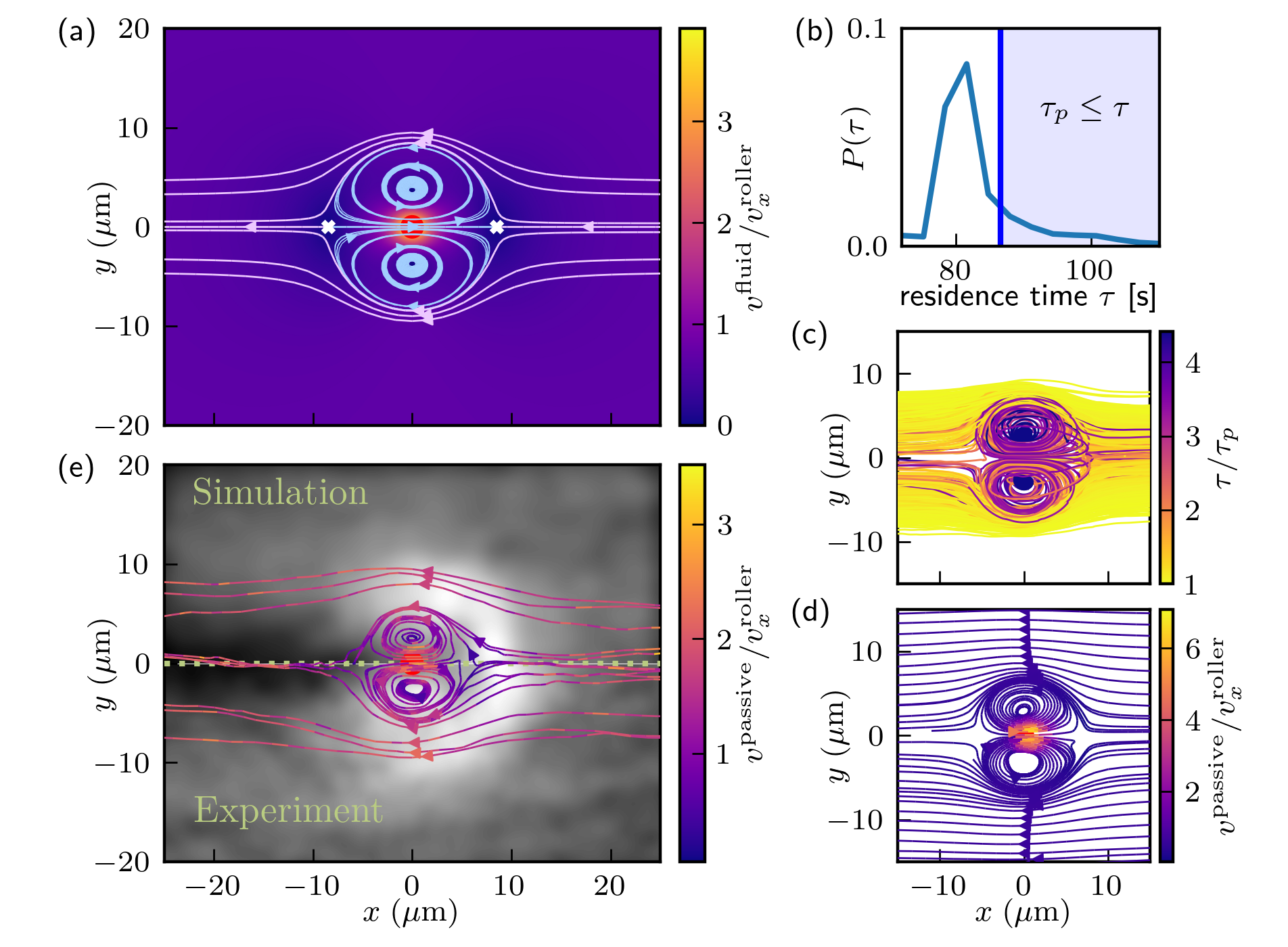}
\caption{\textbf{The emergent pattern results from hydrodynamic interactions around a microroller.} Fluid streamlines in the micoroller's frame produced from a microroller (orange circle) at $z = 3  h_\mathrm{r}$ (a). The fluid velocity is normalized by the microroller's translational speed. There are two characteristic sets of streamlines around the microroller, recirculating (blue) and bypassing (purple) streamlines. Additionally, we observe the presence of stagnation points (white x-crosses) in the front and back of the microroller. In a suspension of passive particles at $T = 0 \, \mathrm{K}$, we extract passive particles that have residence times $\tau$ larger than background particles $\tau_p$ (b) and plot their trajectories (c). These trajectories are confined to the recirculating and bypassing regions. Using the average passive particle velocity profile in (d), we determined that particles in the recirculating region have long residence because they are trapped around the microroller. Meanwhile, particles in the bypassing region persist around the microroller due to their curved trajectories around the microroller. Finally, using simulation data of suspensions at $\mathrm{T} = 293 \, \mathrm{K}$, we show the passive particle streamlines around the microroller and overlay them on the simulation  $\left< \rho_{\mathrm{passive}}\left( \mathbf{r} \right) \right>_t$  (top) and experimental emergent pattern (bottom)(e). This shows that the recirculating and bypassing streamlines span the emergent pattern.     
}
\label{fig:xy_mechanism}
\end{figure*}

As this is an advection-dominated system, we focus on the streamlines generated by the microroller to explain the formation of the emergent pattern. While these streamlines are three dimensional in nature we will show that it is sufficient to focus on streamlines in the $xy$ plane to understand the formation of the emergent pattern. Moreover, the experimental measurements only capture the pattern extent in the $xy$ plane.

We begin by calculating the flow velocity around the rotating microroller in the frame of the microroller, see Fig. \ref{fig:xy_mechanism}(a). In the microroller frame, the microroller is stationary while the passive particles are the mobile species. In the flow profile in the microroller frame, two different sets of streamlines are observed in the fluid in the vicinity of the microroller: (1) a set that surrounds and recirculates around a pair of axial symmetric vortices alongside the microroller (light blue streamlines in Fig.~\ref{fig:xy_mechanism}(a)), and (2) a set that bounds and bypasses the recirculating region (light purple streamlines in Fig.~\ref{fig:xy_mechanism}(a)). An additional feature in the flow field is the appearance of two axially symmetric stagnation points in the front and back of the microroller. These stagnation points are saddle points, that is the fluid flow is convergent along one direction and divergent in the orthogonal direction. In our system, the microroller is always driven in the $+ \hat{x}$ direction; this breaks the symmetry of the stagnation points. The front stagnation point focuses fluid along the $x$ axis and ejects fluid in the $+ \hat{y}$ and $ -\hat{y}$ directions. Meanwhile, the opposite is true for the stagnation point behind the microroller; fluid is focused through streamlines in the $y$ axis and expelled in the $+ \hat{x}$ and $ -\hat{x}$ directions. 


To understand how a pattern emerges in the passive suspension, we first consider how a single passive particle interacts with the flow field generated by the microroller. \comment{To simplify our analysis, we assume that the passive particle does not perturb the microroller streamlines even though they are of finite size. We believe this is reasonable considering the dimensions of the emergent pattern as the distance between the microroller and the passive particles are a distance away where lubrication is negligible.} We begin our analysis by considering a passive particle which remains in the plane of the $xy$ streamlines (above the microroller) as seen in Fig. \ref{fig:xy_mechanism}(a);
a single passive particle approaching the microroller from the right and located around $y=0$  will encounter the front stagnation point. Any slight perturbation will displace the particle in the $+ \hat{y}$ or $ -\hat{y}$ directions and direct the particle into the bypassing streamlines bounding the recirculating region. The particle would then be transported to the back stagnation point where once again a slight perturbation can either push the particle into the recirculating region or eject it in the $- \hat{x}$ direction. Sources of perturbations in the experimental system are thermal fluctuations and near field (lubrication) interactions from other particles. In dense passive particle suspension, near field interactions are prominent due to the proximity of particles in space. 
For equal sized spheres, this occurs when the distance between a pair of particles is equal to or less than two particle diameters. This interaction is largely associated to the squeezing of fluid out from between the narrowly separated particles. When two particles are close enough to each other in the stagnation region, where the fluid flow speed vanishes, near field interactions will enable these particles to travel across streamlines away from the stagnation point and enter either the recirculating or bypassing streamlines.

To explain the emergent pattern observed in experiments, we note that $\left< \rho_{\mathrm{passive}}\left( \mathbf{r} \right) \right>_t$ is set by the residence time of the passive particles in the vicinity of the microroller. The emergent pattern arises from the contrast of residence times between the background passive particles and particles that spend more time near the microroller; passive particles that comprise the emergent pattern are those that remain in the vicinity of the microroller for an amount of time $\tau$ greater than the background passive particles. We calculate the background residence time, $\tau_{p} $, which is the maximum time for background passive particles to spend within a square box that envelops the recirculating region of the fluid flow: 
$$\tau_{p} = \frac{L_x}{\left< v_\mathrm{roller} \right> - \sigma_{v_\mathrm{roller}}},$$ 
where $L_x$ is the box length, and $\sigma_{v_\mathrm{roller}}$ is the standard deviation of the microroller's speed in the $\hat{x}$ direction. Interestingly, even in a suspension of passive particles at $\mathrm{T} = 293 \, \mathrm{K}$ we observe deviations in the height of the microroller due to the near field interactions with passive particles. These height fluctuations result in fluctuations in the microroller's velocity. 
Here we choose $L_x = 20 \, \mu \mathrm{m}$ as this is larger than the recirculation zone at this value of the microroller height. We note that the choice of $L_x$ does not matter as long as the extension of the emergent pattern is contained within the box. This is because the variability of residence times only occurs in the regions of non-negligible hydrodynamic interactions near the microroller. 

There are two possible mechanisms for passive particles to have residence times greater than the background time ($\tau > \tau_{p}$): (i) particles traverse the length $L_x$ slower than the microroller, or (ii) particles travel a distance greater than $L_x$ within the square box with area $L_x^2$. Using the passive particle trajectories from our simulations at $\phi_\mathrm{passive} = 0.17$, $ f = 5 \, \mathrm{Hz}$ and, $\mathrm{T} = 0 \, \mathrm{K}$ we identify the set of passive particles where $\tau > \tau_p$. This set corresponds to the tail end of the distribution of residence times $P(\tau)$ as seen in Fig. \ref{fig:xy_mechanism}(b), where the mean of $P(\tau)$ approximately corresponds to $\frac{L_x}{\left< v_\mathrm{roller} \right> }$. In Fig. \ref{fig:xy_mechanism}(c) we plot the set of individual passive particle trajectories with $ \tau \geq \tau_p$ colored by their residence time normalized by $\tau_p$. Recalling that we are analyzing the particle trajectories in the microroller's frame, we observe that the trajectories with $\tau > \tau_p$ clearly replicate the microroller's streamlines as seen in Fig. \ref{fig:xy_mechanism}(a). This shows that near field interactions between passive particles do not qualitatively affect the trajectories expected from the flow field of the microroller. There are two visible regions with contrasting residence times which directly correlate to the two sets of streamlines from Fig. \ref{fig:xy_mechanism}(a), the recirculating and bypassing streamlines. Residence times within the recirculating region are on average 2.5 times greater than those that bypass it. This is further evidenced by Fig. \ref{fig:xy_mechanism}(d) where we have calculated the average spatial velocity profile of the passive particles and observe that passive particles travel at the background speed and sometimes faster than the microroller. There is an additional set of particles with $\tau > \tau_p$; these are the particles whose trajectories correspond to the bypass region near the recirculation zone. As these curved trajectories are longer than the straight trajectories of the background particles, $\tau > \tau_p$ even though these particles move at the same velocity as the background particles. These results suggest that these two regions are responsible for the emergent pattern in experiments. To test this hypothesis, we perform simulations at finite temperature to more accurately replicate experimental conditions. 


In order to produce simulation results that can be quantitatively compared to experiments, we carry out simulations at T $= 293$~K (see Fig.~\ref{fig:roller}). From these simulations we calculate a pattern length of $8 \ \mu$m, similar to the experimental length ($8.4 \pm 0.9 \mu \mathrm{m}$) and identical to the pattern length of suspensions at $\mathrm{T} = 0 \, \mathrm{K}$. This agreement is expected as thermal fluctuations are negligible in our system. To demonstrate the agreement between the experimental results and the Stokesian dynamics simulations we directly compare the pattern found in experiments with that found in the simulations; the top half of Fig. \ref{fig:xy_mechanism}(e) is the passive particle distribution from the simulations, while the lower half is that obtained from experiments. We observe that the passive particle streamlines associated to the recirculating and bypassing region in Fig. \ref{fig:xy_mechanism}(e) overlap with the emergent pattern. Moreover, these streamlines also reflect thermal fluctuations in the positions of the passive particles. Therefore, we have shown the emergent pattern from suspensions at finite temperature are well described by advective flows generated by the microroller. 

In summary, we have established that the emergent pattern reveals regions of non-negligible hydrodynamic interactions. We have characterized this by demonstrating passive particle residence times around the microroller are extended due to the recirculating and bypassing streamlines produced by a microroller. 
Thus far it has been sufficient to only use information from the $xy$ plane to explain the experimental results. This is due to the fact that the patterns obtained from the experiments are calculated from particle locations projected on the $xy$ plane within the depth of field of the microscope. Experimentally, we lose information away from the focal plane, but particle fluctuations from their average height seem to be negligible. Thus, it is sufficient to only use information from the $xy$ flow plane to trace passive particle trajectories and explain the origin of the emergent pattern.

\begin{figure*}
\centering
\includegraphics[width=.98\linewidth]{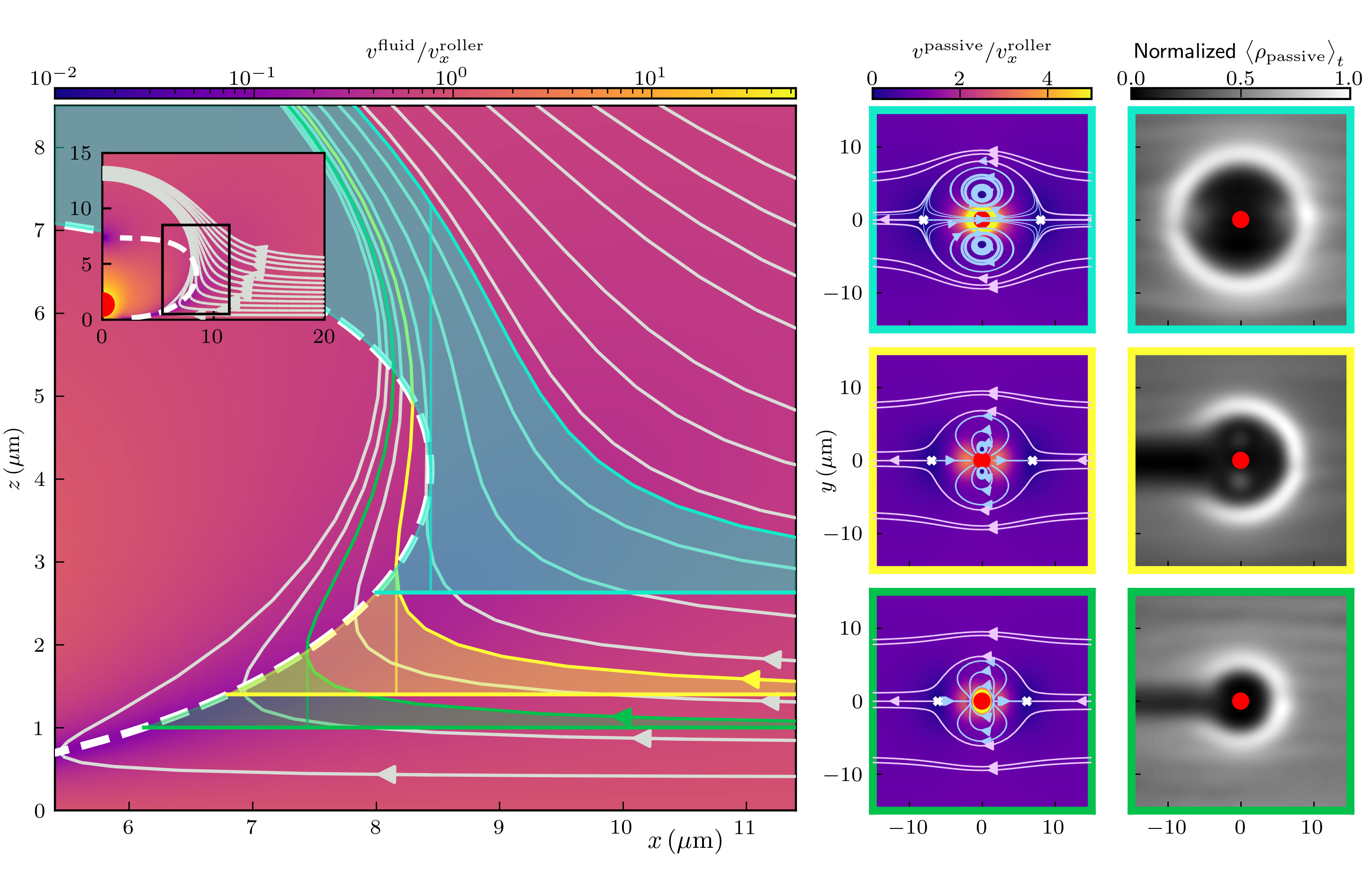}
\caption{\textbf{The microroller's hydrodynamic interactions extend in three dimensions and can be probed by passive particles at different heights.} In the left panel, we plot the $xz$ microroller streamlines at the $(x,0,z)$ plane for a microroller height $ h_\mathrm{roller} = 1.34 \, \mu \mathrm{m}$, and show that not all streamlines (grey curves) intersect the $xy$ stagnation line or saddle line (dashed white curve). Additionally, the saddle line determines the $x$ axis extension of the $xy$ fluid recirculating streamlines at a given height (middle panel). Therefore, different average passive particle heights in suspensions will probe these different recirculating and bypassive streamlines and create different emergent patterns (right panel). We study suspensions with three different particle heights $1.01 \, \mu \mathrm{m}$,  $1.4 \, \mu \mathrm{m}$, and $2.6 \, \mu \mathrm{m}$ which we color code green, yellow and cyan, respectively. In the $xz$ streamlines we bound regions that correspond the the average height of the passive particles and by the $xz$ streamline far from the microroller at the average height of the passive particle. In the middle panel we plot the three different $xy$ streamlines whose spatial extensions are mirrored in suspensions' emergent pattern. We find that the depletion region is present only in the suspensions with particle heights whose bound region near the saddle line is closed.}
\label{fig:xz_mechanism}
\end{figure*}


We now consider how the passive particle height, $h_\mathrm{passive}$, influences the emergent pattern. By modulating the passive particle height, we will sample a different region of the microroller flow field, and, consequently, the emergent pattern will change. To illustrate this, we need to analyze the flow field in three dimensions, and understand how the recirculating and bypassing regions change as a function of passive particle height. Additionally, we need to consider how $xz$ streamlines (Fig. \ref{fig:xz_mechanism} left panel) can lift particles above their average height. This will lead passive particles to interact with different sets of streamlines in the $xy$ plane and lead to emergent patterns with different spatial features (Fig. \ref{fig:xz_mechanism} right panel).  


Previously, we identified the existence of two stagnation points at the front and back of the microroller in the $xy$ plane, see Fig. \ref{fig:xy_mechanism}(a). The region bounded between both stagnation points is the recirculating region of the microroller streamlines in the $xy$ plane and gives rise to the emergent pattern. To understand how the emergent pattern changes with respect to the $z$ axis, it is sufficient to track the front stagnation point as a function of the height from the floor. In Fig. \ref{fig:xz_mechanism}, we show the microroller's $xz$ streamlines at $y = 0$ in the microroller's frame. We superimpose the calculated $xy$ velocity component stagnation line (dashed white line) which correlates to the extension of the recirculating region on the $xy$ plane. Note that the streamlines intersect the stagnation line, meaning the $z$ component velocity is not zero. This curve is not a true stagnation line as not not all velocity components are zero, therefore, it is a saddle line. 

The saddle line's $x$ axis extension varies as a function of $z$. This indicates that the spatial extension of the emergent pattern changes by tuning the height of the passive particles. This is because passive particles will sample different sections of the saddle line. We note that the saddle line is uniquely determined by $h_\mathrm{roller}$. Here we continue to solely focus on passive particle suspensions in water with $h_\mathrm{roller} =  1.34 \, \mu \mathrm{m}$. 

In order to demonstrate the degree of tuning of the emergent pattern's extent with respect to the height of the passive particles, we perform sets of simulations at $\mathrm{T} = 0 \, \mathrm{K}$ with different average passive particle heights $\left < h_\mathrm{passive} \right > $ at constant $\left < h_\mathrm{roller} \right > $. In athermal suspensions, only hydrodynamic forces and the interplay between gravitational and charge repulsion from the floor will change particles' height. We tune $\left < h_\mathrm{passive} \right > $ by varying their excess mass $m_\mathrm{passive}$ with respect to water, and focus on three different $\left < h_\mathrm{passive} \right > $. In suspensions without the presence of a microroller, passive particles have an average height of $1.01 \, \mu \mathrm{m}$,  $1.4 \, \mu \mathrm{m}$, and $2.6 \, \mu \mathrm{m}$ for passive particle excess masses of $10 m_\mathrm{passive}$, $m_\mathrm{passive}$, and with $1/100 m_\mathrm{passive}$, respectively.  

As we have previously stated, suspensions with different $h_\mathrm{passive}$ will form distinct emergent patterns given the curvature of the saddle line. However, we must consider how the $xz$ streamlines in Fig. \ref{fig:xz_mechanism} will impact where passive particles intersect the saddle line. Near the saddle line, $xz$ streamlines can lift particles above their $\left < h_\mathrm{passive} \right >$ and lead to intersect at $z > h_\mathrm{passive}$. In Fig. \ref{fig:xz_mechanism} left panel, we color regions  green, yellow, and cyan, to indicate the range of heights where passive particles will most likely intersect the saddle line for $\left < h_\mathrm{passive} \right > =  \{1.01, 1.4,2.6\} \, \mu \mathrm{m}$, respectively. The lower bound of the colored regions is defined by the particle's $\left < h_\mathrm{passive} \right >$, and the upper bound corresponds to the $xz$ streamline far from the microroller at the respective $\left < h_\mathrm{passive} \right >$. At a distance far from the microroller, $\sim 30 \, \mu \mathrm{m} $, the microroller's hydrodynamic interactions have sufficiently decayed such that the $xz$ streamlines are parallel to one another. This simply represents the background flow which in this reference frame corresponds to the microroller's velocity. 

In general, all $xz$ streamlines curve upwards as approaching the saddle line. This leads to multiple $xz$ streamlines overlapping the different colored regions that passive particles will most likely intersect the saddle line for a given average height. Out of simplicity, we select to intersect the saddle line at the lower bound of this region, the average passive particle height, and plot the $xy$ streamlines around the microroller at this height, see Fig. \ref{fig:xz_mechanism} middle panel (where we show and color the borders of the $xy$ streamlines green, yellow, and cyan to indicate average passive particle heights $\left < h_\mathrm{passive} \right > =  \{1.01, 1.4,2.6\} \, \mu \mathrm{m}$, respectively). As expected, the spatial extension of the recirculating streamlines at different $z$ levels follow the saddle line $x$ values for a given height. In order to show that the emergent patterns obtained from athermal suspensions at different passive particle heights mirror their respective $xz$ streamlines, we plot their $\left < \rho_{\mathrm{passive}}\left( \mathbf{r} \right) \right >_t$ obtained from simulations at $\mathrm{T} = 0 \, \mathrm{K}$
, see Fig. \ref{fig:xz_mechanism} right panel. For all cases, emergent patterns mirror the  $L_\mathrm{pattern}$ of their bypassing streamlines in the $xy$ plane. However, we find a wide variety in the geometry of the emergent pattern. Only suspensions of passive particles with $\left < h_\mathrm{passive} \right > = \{1.01,1.4 \} \, \mu \mathrm{m}$ have a depletion region behind the microroller. Meanwhile, suspensions composed of passive particles with $\left < h_\mathrm{passive} \right >= 2.6 \, \mu \mathrm{m}$ produce an emergent pattern without a depletion region. This difference is produced by the nature of the $xz$ streamlines and their ability to lift particles above their average height. Suspensions that produce emergent patterns with an inverted c-shape structure correspond to passive particle heights in which the majority of the particles will intersect with the saddle line. This can be observed in the colored bounded regions in Fig. \ref{fig:xz_mechanism} left panel, where all $xz$ streamlines in the green and yellow regions are also fully bounded by the saddle line. It is the cyan region which corresponds to passive particles with $\left < h_\mathrm{passive} \right >= 2.6 \, \mu \mathrm{m}$ that contains $xz$ streamlines that miss the saddle line. Thus, not all passive particles intersect with the saddle line and avoid interacting with the recirculating and bypassing streamlines. By not entering the recirculating region, the $xz$  streamlines will transport passive particles above and around the microroller and occupy the region where the depletion region is seen at lower $\left < h_\mathrm{passive} \right >$. 


\comment{Moreover, we briefly explore the origin of the depletion region of the emergent pattern by performing two sets  of simulations at $\mathrm{T} = 0\, \mathrm{K}$. Each set of simulations isolates different interactions between particles. In one case, we remove the hydrodynamic near-field interaction, and the steric interaction that prevents particles from overlapping. In the second set, we leave all near-field interactions intact but simulate mass-less passive particles. We initialize their height at $z = 1.34 \, \mu \mathrm{m}$. Interestingly, the depletion region is still present when near-field interactions are neglected but disappears in suspensions of mass-less passive particles, see Figure S3. This indicates the importance of sedimentation when particles are lifted above their equilibrium height. It has previously been reported\cite{rollers_stability_analysis} that the strength of gravity plays an important role in the dynamics of microroller systems. 

Although this is out of the scope of this paper, we believe that the size of the passive particles should also play a role in the formation of the depletion region. Given that all particles in the suspension are of finite size, they simultaneously sample multiple sets of streamlines and experience shearing. Therefore, a competition between particle size and the local gradients of the streamlines ensues where particles with sizes smaller than the local gradient will mostly be advected while larger particles will be more greatly sheared which will affect their dynamics.
}

Overall, no matter $\left < h_\mathrm{passive} \right > $, all emergent patterns mirror the extension of the recirculating and bypassing regions in the $xy$ plane at their respective height. We should thus expect that for $\left < h_\mathrm{passive} \right > $ above the saddle line no emergent pattern will appear as there would be no region of non-negligible hydrodynamic interactions for the passive particles to sample. Importantly, we have shown that the emergent pattern can be controlled by modifying the height of the passive particles. As we move up from the floor, both recirculating and bypassing regions should increase until they begin to decay as we probe heights further away from the microroller. 



As the saddle line correlates well with the extension of the emergent pattern, we propose $L_\mathrm{pattern}$ to be the distance between the microroller and the saddle line $L_\mathrm{saddle}$. As previously quantified in Fig. \ref{fig:robust}(f), $L_\mathrm{pattern}$ does not change when varying the suspension's passive particle area fraction nor the microroller's velocity, it only changes with the height of the microroller, as does the saddle line. However, to compare $L_\mathrm{saddle}$ and $L_\mathrm{pattern}$, one ambiguity persists, at which height to calculate $L_\mathrm{saddle}$? By analyzing the $xz$ streamlines we have identified a set of bounds for a given $\left < h_\mathrm{passive} \right >$ that delimits the heights passive particles will be driven to by the microroller's $xz$ streamlines. Using these bounds, and knowing both $\left < h_\mathrm{roller} \right >$ and $\left < h_\mathrm{passive} \right >$, we can provide an interval for $L_\mathrm{saddle}$ to compare with $L_\mathrm{pattern}$ obtained from simulations and experiments to show that these two quantities are equivalent. 

\begin{figure}
\centering
\includegraphics[width=0.8\linewidth]{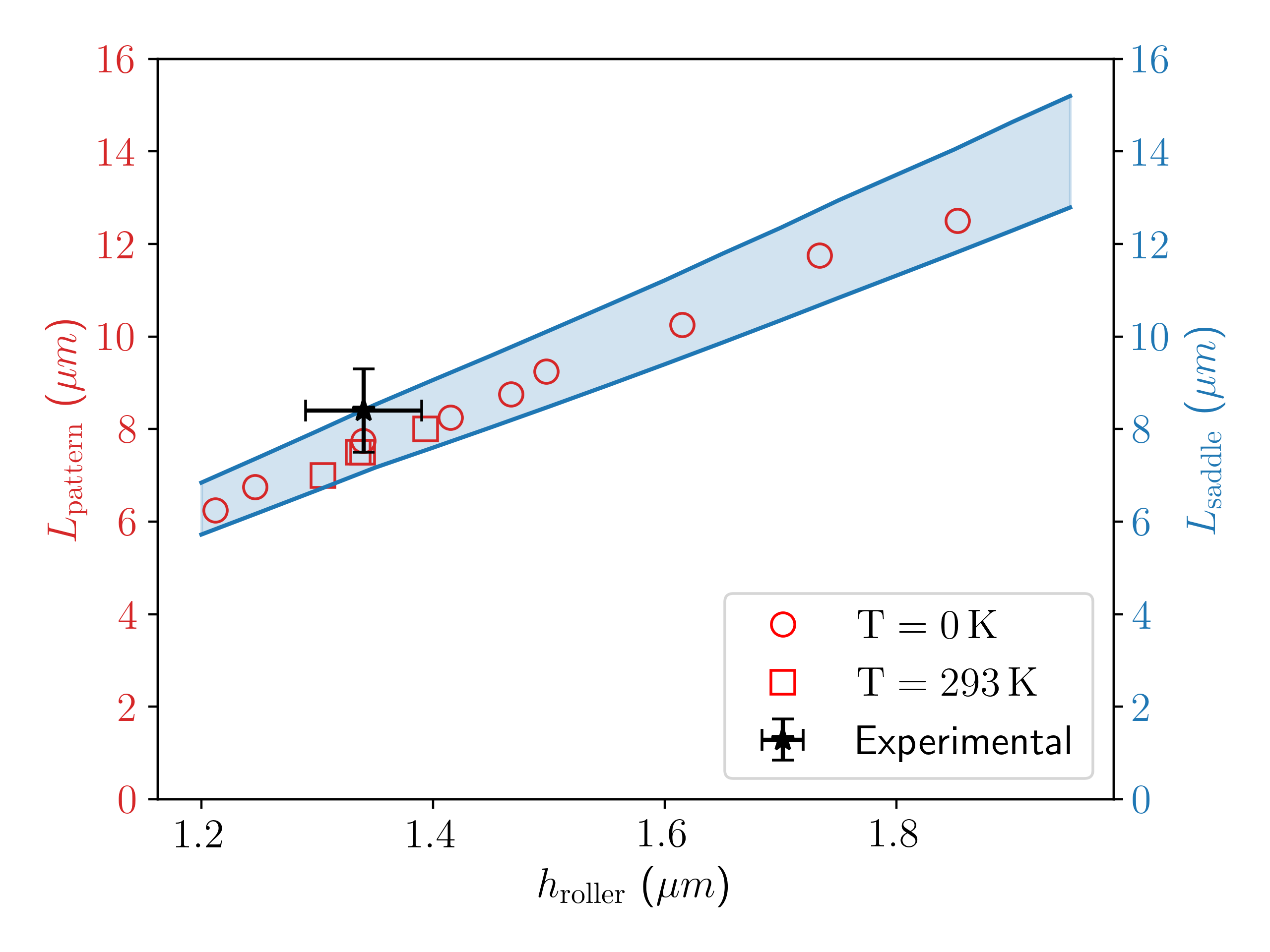}
\caption{\textbf{Tuning the pattern length by changing the microroller height.} We show that the pattern length $L_\mathrm{pattern}$ directly correlates to the distance between the microroller and its front saddle line $L_\mathrm{saddle} \left( z ;\left <  h_\mathrm{roller} \right>)\right)$ (blue region). As previously stated, the saddle line is a function of height, and multiple streamlines in the $xz$ plane intersect a given height which complicates which height to choose to calculate $L_\mathrm{saddle}$. However, we bound the (blue) $L_\mathrm{saddle}$ region by using the bounds determined and shown in Fig. \ref{fig:xz_mechanism}.
\\%
}
\label{fig:L}
\end{figure}

We perform simulations with a microroller at different $\left < h_\mathrm{roller} \right >$ and calculate $L_\mathrm{pattern}$. Furthermore, we compare and parameterize $L_\mathrm{pattern}$ by $L_\mathrm{saddle} \left( z ;\left <  h_\mathrm{roller} \right>)\right)$, where $z$ is the height above the floor. 
As previously discussed, the pattern length has a strong dependency on passive particle height for a given microroller height due to the curvature of the saddle line. Therefore, in simulations at $\mathrm{T} = 0 \mathrm{K}$ we keep the ratio between the passive particle and microroller height approximately constant, $ \left < h_\mathrm{passive} \right > / \left <  h_\mathrm{roller} \right> \approx 1.2 $, to only focus on the effects of $h_\mathrm{roller}$ on the emergent pattern. Meanwhile, the average height of the passive particles in finite temperature suspensions is $ \left < h_\mathrm{passive} \right >  = 2.5 \, \mu \mathrm{m}$. 
In Fig. \ref{fig:L}, we plot $L_\mathrm{pattern}$ and the $L_\mathrm{saddle}$ region, and show that the $L_\mathrm{saddle}$ region is a good descriptor for the emergent pattern obtained from suspensions at $\mathrm{T} = 0 \, \mathrm{K}$, and $\mathrm{T} = 293 \, \mathrm{K}$. Additionally, the extension of the pattern length and $L_\mathrm{saddle}$ decreases as the microroller approaches the no-slip boundary. This is expected, as moving closer to the surface effectively screens the hydrodynamic interactions, reducing their extent. The opposite is true for a microroller further away from the surface. However, as the microroller's height is increased the coupling between the microroller's rotation and translation diminishes until the translation velocity becomes infinitesimally small. In a suspension at finite temperature, thermal fluctuations would then disrupt the pattern created by advective flows. 

\section{Conclusion}
In conclusion, we have demonstrated the ability to use a driven particle to create a large-scale (10 times the particle radius), asymmetrical 3D pattern from a quasi-2D colloidal suspension. The pattern includes an accumulation region with its center being several particle sizes away from the microroller, and a depletion region along the microroller trajectory.
This pattern is created via hydrodynamic interactions, and is unmodified by thermal fluctuations, passive particle area fraction, or driving velocity. 
We show two main pathways to modify the pattern by altering the hydrodynamic interactions between the microroller and the passive particles. This can be done by tuning the height of the passive particles in the suspension with respect to the microroller, or by modifying the height of the microroller with respect to the floor. 
Our analysis of the microroller-driven advective flow that generates the pattern demonstrates that the extension of the emergent pattern is equivalent to the distance between the microroller and the flow's saddle line. 
Thus, modifying the average height of the microroller changes the pattern's size as it changes the features of its fluid velocity profile. Additionally, by modifying the passive particle height in a suspension at constant microroller height, the particles are able to sample other planes of non-negligible hydrodynamic regions and change the extension of the pattern, which demonstrates the three dimensional nature of the microroller's streamlines. 

Our analysis reveals that the pattern scale is determined by equilibrium quantities: the microroller height and the height distribution of the passive particles. Thus, the size scale of the emergent pattern provides an alternative pathway to determine an approximate average passive particle height in a suspension if the $h_\mathrm{roller}$ is known. This analysis is straightforward, is not computationally demanding, and offers a new tool for studying fluid-mediated interactions of driven particles.
If there are weak or transient interaction between the passive particles (for example in a colloidal gel), this pattern formation could be exploited for material restructuring. 
This system also offers an alternative way to do active microrheology \cite{zia_active_2018}. For example, we observe that a microroller moves slower when it is in a colloidal suspension than when it is in a pure fluid. One can thus calculate the effective viscosity of the colloidal suspension by measuring the change of the microroller speed as a function of colloidal volume fraction, and measure density fluctuations by measuring roller velocity fluctuations.
Additionally, similar principles of microroller streamlines can perhaps be used to explain how a mixture of passive particles and biologically active swimmers lead to anomalous transport coefficients of the passive particles via hydrodynamic interactions \cite{mino_enhanced_2011, jepson_enhanced_2013}. 
Finally, we note that when the passive particles enter the recirculating steamlines, they are trapped and move together with the microrollers. Therefore, microrollers that generate these streamlines, or microvortices, have the potential to transport micron-size particles. 

\section*{Materials and Methods}
\subsection*{Experiments}
The passive particles are spherical and made of polystyrene (Bangs laboratory\copyright, FSPP005) with a density of $ 1.06$~g/cm$^3$ and a mean diameter of $= 2.07 \pm 0.15 \, \mu \mathrm{m}$. 
The microrollers in the experiment are described in detail in \citet{sprinkle_driven_2020} and \citet{van_der_wee_simple_2023}. 
The microroller has a mean diameter of $2.1 \pm 0.1 \, \mu \mathrm{m}$ and a permanent magnetic dipole as it is comprised of a hematite cube within a spherical polymer matrix, see Fig.~\ref{fig:roller}(b). The mean density of the microroller is $1.74$~g/cm$^3$.
We clean the passive particles by replacing the solution with DI water for three times. Then we add a small amount of microroller solution to the passive particle solution, and mix the solution with a vortex followed by a sonicator. We withdraw the mixture solution with a capillary tube, and seal the tube entrance with glue. Then, we mount the sample on a microscope, and check the area fraction of passive particles after all particles sediment to the floor. Finally, we apply rotating magnetic fields and record the particle distribution with the fluorescent microscope. The microrollers and the passive particles have different fluorescent wavelength, giving us the ability to separate the two types of particles in two channels.

The system and the mechanism to drive a microroller is described in detail in \citet{sprinkle_driven_2020}. In short, we use two pairs of Helmholtz coils to generate an external rotating magnetic field ($100$~G). The permanent dipole of a driven particle experiences the torque from the external field, causing the driven particle to rotate synchronously with the field. As the microroller is near the floor, the flow generated by the rotating driven particle becomes asymmetrical due to the non-slip boundary of the floor, which causes the microroller to translate.
We trace the location of the microrollers using Python and the package \href{https://soft-matter.github.io/trackpy/v0.6.1/}{Trackpy}, which we use to generate a sequence of images around the microroller. We then use the position of the microroller to shift the coordinates of all images to the microroller frame.

\subsection*{Simulations}

As we have shown, the predominant interaction between a microroller and passive particles in suspension are hydrodynamic in nature. To correctly quantify these interactions we simulate these systems using lubricated corrected Brownian dynamics \cite{sprinkle_driven_2020}. In this method the position and orientation of a particle $\mathbf{q}_1 = \{ \mathbf{x}, \mathbf{\theta} \}$ are evolved by 
\begin{equation}
    \frac{\mathrm{d} \mathbf{Q}}{\mathrm{d} t} = \overline{\mathbf{M}} \mathbf{F} + k_\mathrm{B} \mathrm{T} \partial_\mathbf{Q} \cdot \overline{\mathbf{M}} + \sqrt{2 k_\mathrm{B} \mathrm{T} \overline{\mathbf{M}}} \, \mathcal{\mathbf{W}}(t)
\label{eq:BD}
\end{equation}
where $\mathbf{Q} = \left [ \mathbf{q}_1, \mathbf{q}_2, \dots, \mathbf{q}_n \right ]$ is the vector containing the the individual positions and orientations of all particles. Here, pairwise hydrodynamic interactions between particles are determined by their configuration in space and contained in the lubrication corrected mobility matrix $\overline{\mathbf{M}} \left ( \mathbf{Q} \right )$. The magnitude of these interactions are weighted by deterministic properties governed by external forces $\mathbf{f}$ and torques $\mathbf{\tau}$ acting on particles in solution, and stochastic properties arising from the presence of thermal fluctuations. The first term with respect to the right of Eq. \ref{eq:BD} details its deterministic character, here $\overline{\mathbf{M}}$ is weighted by the vector comprised of individual external forces and torques on impinging on all particles $\mathbf{F} = \left [ \mathbf{f}_1, \mathbf{\tau}_1, \dots, \mathbf{f}_n, \mathbf{\tau}_2, \right ]$. The second and third terms of the equation deal with the thermal drift, and random walk nature of the of the thermal fluctuations, respectively. Here, $k_\mathrm{B}$ denotes the Boltzmann constant, $\mathrm{T}$ indicates the solvent temperature, and $\mathcal{\mathbf{W}}(t)$ represents a Wiener process or a collection of independent white noise processes essential for the generation of Brownian velocities. Given that this is a stochastic differential equation, we temporally integrate this equation using a stochastic scheme, specifically the ‘Stochastic Trapezoidal Split’ (STS) scheme \cite{sprinkle_driven_2020}. In this paper we perform simulations of Eq. \ref{eq:BD} evolved by the STS scheme using a publicly accessible code found on github at \href{https://github.com/stochasticHydroTools/RigidMultiblobsWall}{RigidMultiblobsWall}. More details including the accuracy of this scheme, and pre-conditioners employed in the solution for $\overline{\mathbf{M}}^{1/2}$ can be found in \citet{sprinkle_driven_2020}.  

To simulate suspensions of passive particles with a microroller, we model passive particles and the microroller as spherical rigid particles with radii $R_\mathrm{passive} = 1.0 \, \mu \mathrm{m}$ , and $R_\mathrm{roller} = 1.0 \, \mu \mathrm{m}$, respectively. The particles are immersed in water at $T = 293.15 \, \mathrm{K}$, and they have a buoyant mass of $m_\mathrm{passive} = 2.5 \, \times \, 10^{-16} \, \mathrm{kg} $, and $m_\mathrm{roller} = 3.1 \, \times \, 10^{-15} \, \mathrm{kg}$, respectively. We perform three dimensional simulations with an initial condition set by fixing the area fraction of passive particles and randomly populating passive particles in a two dimensional strip of $250 \, \mu \mathrm{m}$ in length and $75 \, \mu \mathrm{m}$ width at $z = 1.5 \, \mu \mathrm{m}$. Finally, we place a non-rotating microroller to the left of the strip, and equilibrate particle positions by evolving the system for approximately $60 \, \mathrm{s}$, after which we rotate the microroller at constant angular velocity using the algorithm detailed in \citet{sprinkle_driven_2020}. For all instances of the simulations we use a time step of $\Delta t = \, 0.05 \, \mathrm{s}$. 

In this paper, all particles experience gravitational forces given their excess mass, and electrostatic repulsion with the lower surface. We model the electrostatic repulsion using the Yukawa type potential:
\begin{equation}
    U (h) = \begin{cases}
    \epsilon \exp{\left ( \frac{R - h}{\kappa}  \right)} & \text{if}\; h > R\\
    \epsilon \left ( 1 - \frac{R - h}{\kappa} \right ) & \text{if} \; h < R
    \end{cases}
    ,
    \label{eq:yukawa}
\end{equation}
where $h$ is the particle center to floor distance, and $R$ is the radius of the respective particle. For the microroller, we set the magnitude of the potential $\epsilon$, and its screening factor $\kappa$ that simultaneously matches the height that replicates its experimentally obtained velocity profile $\left ( h_\mathrm{roller} = 1.34 \, \mu \mathrm{m} \right )$, and its measured diffusion coefficient, $D_\mathrm{roller} = 0.15 \, \mu \mathrm{m}^2/\mathrm{s}$. For the passive particles, we assume the same $\kappa$ as the microroller, and instead fit $\epsilon$ to match its experimentally obtained diffusion coefficient, $D_\mathrm{roller} = 0.015 \, \mu \mathrm{m}^2/\mathrm{s}$. The list of parameters used in the simulations can be found tabulated in the SI. Additionally, as the suspension is located above a no-slip wall, hydrodynamic interactions are calculated using Blake's solution of the Green's function solution to the Stokes equation above a no-slip wall generalized for spherical rigid particles \cite{Blake1971,Swan_no_slip}. However, this Green's function only correctly describes far field hydrodynamic interactions between all pairs of surfaces, particle-particle and particle-wall. To include near field hydrodynamic interactions related to the squeezing of fluid between pairs of surfaces we use the previously mentioned lubrication corrected mobility matrix $\overline{\mathbf{M}}$ detailed in \citet{sprinkle_driven_2020}. Moreover, we use two different cutoffs that determines at which distance between surfaces at which to calculate either near field hydrodynamic interactions or far field hydrodynamic interactions. For particle-particle interactions the cutoff distance is $r \leq 5 \, \mu \mathrm{m}$ where for distances greater than $5 \, \mu \mathrm{m}$ we simply calculate interactions with the Green's function solution. Meanwhile, we calculate near field hydrodynamic interactions for particle-wall interactions for any distance above the wall. For more information about the implementation of the resistance scalars for near field interactions can be found in \citet{sprinkle_driven_2020}. We additionally complement near field hydrodynamic interactions with a short-ranged steric repulsion potential $U_\mathrm{cut} \left (r \right)$ with the Yukawa type potential of Eq. \ref{eq:yukawa} for particle-particle and particle-wall interactions. In the case for particle-particle interactions we substitute $R = 2R \,(1 - \delta_\mathrm{cut})$, while $R= R \, (1 - \delta_\mathrm{cut})$ for particle-wall interactions, where $\delta_\mathrm{cut} = 10^{-3} \, \mu \mathrm{m}$. The complete set of parameters used in $U_\mathrm{cut} \left (r \right)$ are also tabulated in the SI.

Additionally, we calculate $\left < \rho_\mathrm{passive} \right >_t$ averaging over at least 10 different simulation runs using a 30 $\mu \mathrm{m} \, \times \, 30  \mu \mathrm{m}$ mesh with bin width $\Delta L = 0.25 \mu \mathrm{m}$. We choose to average over frames $0.1 \, \mathrm{s}$ apart and where the microroller is $L_x \in [20,230] \, \mu \mathrm{m}$. Under these bounds the microroller is within the region of passive particles with a given area fraction $\phi$. We also use this implementation to avoid averaging over regions outside the bounds of the suspension which affects the formation and dimensional features of the emergent pattern. After calculating $\left < \rho_\mathrm{passive} \right >_t$ we use Gaussian smoothing with a variance the size of the passive particle radius to include particle areal size effects. This allows comparison between $\left < \rho_\mathrm{passive} \right >_t$ and experimental emergent patterns now that $\left < \rho_\mathrm{passive} \right >_t$ contains information of the passive particle size and loosens the constraints on the distributions obtained by using a homogeneous binning mesh of $0.25, \mu \mathrm{m}$. We calculate all velocity distributions of a microroller also using the Blake's solution to the Green's function of a stokeslet above a no-slip wall generalized for a spherical particle \cite{Swan_no_slip}.

\section*{Conflicts of interest}
There are no conflicts to declare.

\section*{Acknowledgements}
We thank Brennan Sprinkle for fruitful discussions concerning Brownian motion of low density particles, and Bhargav Rallabandi for discussions on the P\'eclet number and tracers. H.L.-R. acknowledges support from a MRSEC-funded Graduate Research Fellowship, (DMR-2011854). This work was primarily supported by the University of Chicago Materials Research Science and Engineering Center, which is funded by National Science Foundation under award number DMR-2011854.



\balance


\bibliography{pattern} 
\bibliographystyle{rsc} 

\end{document}